\documentclass[twocolumn,aps,superscriptaddress,english]{revtex4-2}
\usepackage[colorlinks=true,allcolors=blue]{hyperref}
\usepackage{array}
\usepackage{amsmath}
\usepackage{amssymb}
\usepackage{graphicx}
\usepackage{babel}
\usepackage{braket}
\usepackage{mathrsfs}
\usepackage{amsfonts}
\usepackage{epstopdf}
\usepackage{multirow}
\usepackage{color}
\usepackage{bm}
\usepackage{amsthm}

\usepackage{xcolor}
\usepackage{tikz,fp}
\usepackage{tikz-cd}
\usepackage{adjustbox}
\usepackage{enumitem}
\usepackage{makecell}
\usetikzlibrary{arrows}
\usetikzlibrary{intersections}
\usetikzlibrary{shapes.geometric}
\usetikzlibrary{decorations.pathmorphing, patterns,shapes}
\usetikzlibrary{decorations.markings}
\usetikzlibrary{snakes}
\tikzset{
  mid arrow/.style={postaction={decorate,decoration={
        markings,
        mark=at position .575 with {\arrow[#1]{stealth}}
      }}},
  near arrow/.style={postaction={decorate,decoration={
        markings,
        mark=at position .275 with {\arrow[#1]{stealth}}
      }}},
   far arrow/.style={postaction={decorate,decoration={
        markings,
        mark=at position .800 with {\arrow[#1]{stealth}}
      }}},
}
\tikzset{snake it/.style={decorate, decoration=snake}}

\usetikzlibrary{arrows}
\usetikzlibrary{intersections}
\usetikzlibrary{shapes.geometric}
\usetikzlibrary{decorations.pathmorphing, patterns,shapes,fixedpointarithmetic}
\usetikzlibrary{decorations.markings}
\usetikzlibrary{decorations.shapes}

\newcommand{\dif}[1]{\operatorname{d}\!#1\,} %use for differential operator
\newcommand{\ee}{\mathrm{e}} %use for exponential
\newcommand{\ii}{\mathrm{i}} %use for imaginary number
\newcommand{\Id}{\mathbb{I}} %use for identity matrix
\newcommand{\Tr}{\mathrm{Tr}}
\newcommand{\TA}{\widetilde{\mathcal{A}}} 
 
\newcommand{\TC}{\widetilde{\mathcal{C}}}

\renewcommand{\AA}{\mathcal{A}} 
\newcommand{\BB}{\mathcal{B}} 
\newcommand{\CC}{\mathcal{C}} 
\newcommand{\DD}{\mathcal{D}} 
\newcommand{\RR}{\mathcal{R}}

\newcommand{\dist}{{\rm dist}}
\newcommand{\StirII}[2]{\left\{ {#1 \atop #2} \right\}}

\renewcommand{\hat}[1]{\widehat{#1}}

\renewcommand{\ket}[1]{\lvert#1\rangle}
\renewcommand{\bra}[1]{\langle#1\rvert}
\newcommand{\lrangle}[1]{\left\langle#1\right\rangle}

\renewcommand{\vec}[1]{\boldsymbol{\mathbf{#1}}}

\makeatletter

\newcommand*{\wideboxed}[1]{\setlength{\fboxsep}{1ex}%
  \fbox{\m@th$\displaystyle#1$}}
\makeatother

\makeatletter
\def\maketitle{
\@author@finish
\title@column\titleblock@produce
\suppressfloats[t]
}
\makeatother
\begin{document}

\title{Entanglement Measure Response to Modular Flow and Chiral Topological Phases}
\author{Yunlong Zang}
\email{zangyunlong22@mails.ucas.ac.cn}
\affiliation{Kavli Institute for Theoretical Sciences,\\ University of Chinese Academy of Sciences, Beijing 100190, China}
\date{\today}

\begin{abstract}
Recent years have witnessed significant progress in the entanglement-based characterization of quantum phases of matter. The primary objects of interest are the reduced density matrix and its associated entanglement Hamiltonian. As intrinsic properties of a quantum state, these quantities theoretically determine all experimentally accessible local observables. In this work, we investigate the response of two entanglement measures to the real-time dynamics driven by the entanglement Hamiltonian—a process known as modular flow. 
We demonstrate that our results can be unified into a single generating function, $\langle\rho_{AB}^\alpha \mathrm{e}^{\lambda {Q}_{AB}}\mathrm{e}^{\mu {Q}_{BC}}\rho_{BC}^\beta\rangle$. This function is of independent interest as it represents a generalization of the recently proposed R\'{e}nyi modular commutator.
 In appropriate limits, this function yields the response of R\'{e}nyi entropy and its charged version, which we find to be uniquely determined by chiral topological invariants, specifically the chiral central charge and the Hall conductance. Our analytical findings are validated through two independent approaches: (i) free fermion systems using the real-space Chern number formula, and (ii) an effective field theory treatment that regularizes the entanglement cut via chiral conformal field theory. Both methods yield consistent results.
\end{abstract}

\maketitle

\section{Introduction}

The efficient characterization of topological phases remains a central challenge in condensed matter physics, as these phases defy description by traditional local order parameters. In two-dimensional (2D) gapped systems, a hallmark example is topological entanglement entropy~\cite{Kitaev2006TEE,Levin2006detecting}, where the constant term is uniquely determined by the quantum dimensions of anyons. This discovery has paved the way for characterizing quantum phases through entanglement-related quantities.

At the heart of these entanglement measures lies the reduced density matrix (RDM). The RDM of a subsystem encapsulates all local observables, which constitute the primary data accessible in experiments. This suggests that robust detection methods should be designed around the RDM. In recent years, this approach has successfully produced various real-space observables~\footnote{See, for example Ref.~\cite{Tu2013Momentum,Shiozaki2017,Shiozaki2018, Cian2021,liu2024anyon,h53w-64dy}} for 2D topological phases. Theoretically, the RDM of a local patch is conjectured to be able to derive universal properties of a phase, as demonstrated by the entanglement bootstrap program~\cite{SHI2020} and Kitaev’s proposal for the local definition of quantum phases~\cite{KitaevLocal}.

\begin{figure}[t]
\centering
 \adjincludegraphics[scale=1,valign=c]{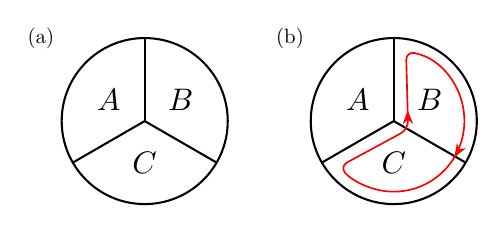}
\caption{The geometric setting. (a): A large but finite disk $ABC$ is tripartited. (b): Modular flow causes chiral motion of entanglement modes (red lines) in chiral topological phases.}
\label{fig:ABCsetting}
\end{figure}

A prominent realization of this framework is the modular commutator proposed in Ref.~\cite{Kim2022chiral,Kim2022modular,Fan2022from,Fan2023Extracting}. Consider a 2D quantum state defined on an infinite plane, with a large, disk-shaped subregion partitioned into three adjacent sections, $A$, $B$, and $C$, as illustrated in Fig.~\ref{fig:ABCsetting}(a). The core relation is given by:
\begin{equation}
\left.\frac{\dif{}}{\dif{s}} \right|_{s=0} S_{AB}(\rho_{ABC}(s)) =\ii\lrangle{[K_{AB},K_{BC}]} =\frac{\pi}{3}c_{-}.
\label{eq:mc}
\end{equation}
To clarify our notation: for any subregion $R$, the RDM is denoted by $\rho_R$ with entanglement Hamiltonian $K_R = -\log \rho_R$, and the entanglement entropy is $S_R(\rho)=\Tr(\rho K_R)$. Here, $\rho_{ABC}(s)$ represents the RDM $\rho_{ABC}$ evolves under the  modular flow generated by $K_{BC}$:
\begin{equation}
\rho_{ABC}(s) = \ee^{\ii sK_{BC}} \rho_{ABC} \ee^{-\ii s K_{BC}}
\label{eq:rhoABCS}
\end{equation}
where $s$ is the modular parameter. Ultimately, by evaluating the derivative of the entanglement entropy of region $AB$ with respect to this modular evolution, we obtain a quantity directly proportional to the chiral central charge $c_{-}$. Its validity has been rigorously proven for free fermions~\cite{Fan2023generalized}. For interacting systems, this quantity is supported by arguments based on bulk-edge correspondence and conformal field theory calculation~\cite{Fan2022from}, though a rigorous proof remains elusive.

Intuitively, the modular commutator reflects the response of the entanglement entropy to modular flow, where the specific tripartite geometry plays a critical role. As illustrated in Fig.~\ref{fig:ABCsetting}(b), the modular flow generated by $K_{BC}$ induces a motion of the entanglement degrees of freedom (d.o.f.s) along the boundary. This flow essentially redistributes the entanglement between region $AB$ and its complement. By accounting for the net flux of these d.o.f.s across the entanglement cuts, one can quantify the total gain or loss of entanglement for region $AB$. To formalize this heuristic picture, we assume that the density of these area-law entanglement d.o.f.s is directly proportional to the total central charge $c_L +c_R$ and then the net flux proportional to their difference $c_-=c_L-c_R$~\cite{Fan2022from}\footnote{This argument implies that the density of entanglement degrees of freedom, $\sigma$, is inversely proportional to its rate of change $v$ under modular flow. Consequently, the product $\sigma v$ remains dimensionless.}.

In this paper, we generalize the aforementioned heuristic picture to broader entanglement measures, with a primary focus on Rényi entropy and its charged counterpart. We demonstrate that these quantities exhibit a universal response to modular flow, which is remarkably UV-independent. This development parallels recent investigations into the R\'{e}nyi modular commutator~\cite{sheffer2025probing,gass2025renyilike}. Specifically, our proposed generating function represents a direct generalization by simultaneously incorporating the density operator and the $U(1)$ charge operator. Our formal derivation adopts and extends the logical framework established in Refs.~\cite{sheffer2025probing,gass2025renyilike}.

\section{Main Results}
\emph{Setup.---}
 In this paper, we consider 2D gapped systems possessing a global $U(1)$ symmetry generated by $Q$. Such systems have two chiral topological invariants: the chiral central charge $c_{-}$ and the Hall conductance $\sigma_{xy}$. For a large tripartite subregion $ABC$ topologically equivalent to the configuration in Fig.~\ref{fig:ABCsetting}(a), we find that the generating function $\Omega_{\alpha,\beta,\lambda,\mu}$ satisfies the following universal relation:
\begin{equation}
\begin{aligned}
&\quad\quad\quad\quad\Omega_{\alpha,\beta,\lambda,\mu}=\lrangle{\rho_{AB}^\alpha\ee^{\lambda{Q}_{AB}}\ee^{\mu{Q}_{BC}}\rho_{BC}^\beta}\\
    & \arg \Omega_{\alpha,\beta,\lambda,\mu}=-\frac{\pi c_{-}}{12} q(\alpha,\beta)+\frac{\sigma_{xy}}{2}\lambda^2s(\alpha,\beta)+\frac{\sigma_{xy}}{2}\mu^2s(\beta,\alpha)\\
    &q(\alpha,\beta) =\frac{\alpha}{\alpha+1}+\frac{\beta}{\beta+1}-\frac{\alpha+\beta}{\alpha+\beta+1} \\
    &s(\alpha,\beta) =\frac{\alpha+\beta}{\alpha+\beta+1}-\frac{\alpha}{\alpha+1}
\end{aligned}
\label{eq:def_omega}
\end{equation}
where $Q_R$ is subsystem charge of region $R$ and $\arg$ denotes the complex phase, i.e., $\arg(r e^{\mathrm{i}\theta})=\theta$. This implies that only the phase of $\Omega_{\alpha,\beta,\lambda,\mu}$ captures universal topological information. In this definition, $\alpha$ and $\beta$ serve as replica indices that can be analytically continued to any positive real value. Conversely, $\lambda$ and $\mu$ are real parameters, meaning the associated operators do not correspond to unitary disorder operators. The relation in Eq.~\eqref{eq:def_omega} remains valid provided the dimensionless parameters $|\lambda|$ and $|\mu|$ are sufficiently small compared to the ratio $\ell_{ABC}/\xi$, where $\ell_{ABC}$ is the characteristic linear size of the region and $\xi$ is the correlation length.

\emph{Application to modular flow.---}
The generating function directly yields the universal response of various entanglement measures to modular flow. For any region $R$, we define its $\alpha$-R\'{e}nyi entropy and its charged counterpart as:
\begin{equation}
    S_R^{(\alpha)} = \frac{1}{1-\alpha}\log\Tr \rho_R^\alpha, \quad S_R^{Q(\alpha)} = \frac{1}{1-\alpha}\log\Tr \ee^{\lambda Q_R}\rho_R^\alpha.
\end{equation}
By subjecting the RDM $\rho_{ABC}$ to the modular flow $\rho_{ABC}(s)$ as defined in Eq.~\eqref{eq:rhoABCS}, we find:
\begin{equation}
\begin{aligned}
     &\left.\frac{\dif{}}{\dif{s}} \right|_{s=0} S_{AB}^{(\alpha)}(\rho_{ABC}(s)) = \frac{\pi c_{-}}{6} \frac{\alpha+1}{\alpha} \\
     &\left.\frac{\dif{}}{\dif{s}} \right|_{s=0} S_{AB}^{Q(\alpha)}(\rho_{ABC}(s)) = \frac{\pi c_{-}}{6} \frac{\alpha+1}{\alpha} -\sigma_{xy}\frac{\lambda^2}{\alpha(\alpha-1)}\\
\end{aligned}
\end{equation}
Remarkably, both expressions are governed solely by the replica index $\alpha$ and the chiral topological invariants, remaining entirely independent of UV details. The first line recovers the modular commutator result in Eq.~\eqref{eq:mc} in the von Neumann limit ($\alpha \to 1$). In contrast, the second line exhibits a singularity, indicating that the $\alpha \to 1$ limit cannot be taken naively for the charged version.

In the following, we first prove Eq.~\eqref{eq:def_omega} for free fermion systems in Sec.~\ref{Sec:freefermion}. Subsequently, Sec.~\ref{Sec:app} details the derivation of the universal response of various entanglement measures to modular flow. Furthermore, we provide an alternative proof in the Appendix~\ref{Sec:field_theory} based on effective field theory. Additional technical details and supporting derivations are included in Appendix~\ref{Sec:details}.

\section{Free fermion case}
\label{Sec:freefermion}
\begin{figure}[]
\centering
\adjincludegraphics[scale=1,valign=c]{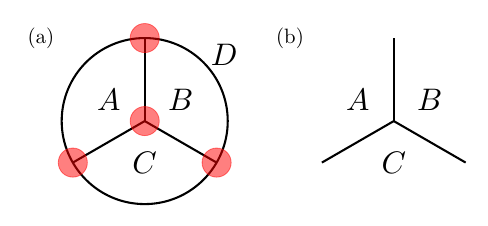}
\caption{Free fermion systems. (a): A large but finite disk ABC is tripartited, with its complementary region labeled by D. Four red shaded regions are triple contact points. (b): An infinite plane is tripartited. In both cases, only the topology is essential and a specific shape does not matter.}
\label{fig:freefermion}
\end{figure}
We now present a proof for free fermion systems. In this regime, all observables are fully characterized by the two-point correlation function~\cite{Peschel2003}, also referred to as the \emph{spectral projector} $P$, with matrix elements $P_{jk}=\lrangle{c_k^\dagger c_j}$.
For gapped Hamiltonians, $P$ satisfies the quasi-diagonal property (exponential decay): 
\begin{equation}
|P_{jk}|\le c \ee^{-|j-k|/\xi}.
\end{equation}
We further define the spatial projection matrix $\RR$ for a generic subregion $R$, with elements $(\RR)_{jk} = \delta_{jk}$ if $j,k \in R$ and $0$ otherwise. Consequently, the correlation matrix restricted to region $R$ is given by $P_R \equiv \RR P \RR$. Using the properties of Gaussian states~\cite{Klich2003}, the generating function can be expressed as a determinant:
\begin{equation}
\begin{aligned}
      &\quad\quad\quad\quad\Omega_{\alpha,\beta,\lambda,\mu}=\det M\\
      &M= \Id-P+P(\ee^\lambda P_{AB}^\alpha+P_{CD}^\alpha)(\ee^\mu P_{BC}^\beta+P_{AD}^\beta)P,\\
      \label{eq:def_M}
\end{aligned}
\end{equation}
where $\Id$ is the identity matrix, and we have assumed the total system is partitioned into $ABCD$.
An explicit derivation of Eq.~\eqref{eq:def_M} can be found in Appendix~\ref{Sec:details}.

In free fermion systems with charge conservation, the chiral topological invariant is given by $\nu(P)=c_{-}=2\pi\sigma_{xy}$, where $\nu(P)\in \mathbb{Z}$ denotes the \emph{real-space Chern number} determined by the spectral projector $P$. The original formulation of $\nu(P)$ was introduced by Kitaev~\cite{KITAEV2006}, while a higher-order generalization was recently proposed in Ref.~\cite{Fan2023generalized}. In this work, we employ a refined version of this invariant~\cite{gass2025renyilike}:
\begin{equation} 
 \Tr[f(PCP),g(PAP)]=\ii\frac{\nu(P)}{2\pi}\int_0^1 [f(t)-f(0)]\partial_t g(1-t) \dif t,
\label{eq:rschern}
\end{equation}
where functions $f$ and $g$ are analytic at $0$. Note that Eq.~\eqref{eq:rschern} is strictly valid only for the infinite plane geometry depicted in Fig.~\ref{fig:freefermion}(b).

Following the methodology of Ref.~\cite{gass2025renyilike}, we demonstrate that the dominant contributions to Eq.~\eqref{eq:def_M} are localized near the four triple junctions (shaded regions in Fig.~\ref{fig:freefermion}(a)). In the vicinity of each junction, we can take the thermodynamic limit—effectively treating the local partition as covering the entire space ($\AA+\BB+\CC=\Id$). Under this approximation, Eq.~\eqref{eq:rschern} becomes applicable, allowing us to explicitly evaluate the contributions.

\subsection{Reduction to triple contact points}
In this subsection, we analyze the dominant contributions to Eq.~\eqref{eq:def_M} by exploiting the quasi-diagonal property of $P$. 
\subsubsection{Bulk and interface contributions}
We first consider contributions from the deep interior of a single region. Let ${B}_0$ denote such a region; specifically, $j\in {B}_0$ if $\dist(j, {ACD})$ is sufficiently large compared to the correlation length $\xi$. Restricted to this subspace, matrix elements $P_{jk}$ are exponentially suppressed if $k\notin {B}$. Consequently, we find:
\begin{equation}
\begin{aligned}
    {\BB_0}M {\BB_0}&\approx  {\BB_0}(\Id-P+\ee^{\lambda+\mu} P P_B^{\alpha+\beta}P) {\BB_0} \\
    &\approx  {\BB_0}(\Id-P+\ee^{\lambda+\mu} P) {\BB_0}.
\end{aligned}
\end{equation}
where '$\approx$' indicates that the error is of order $\ee^{-\dist({B}_0,{ACD})/\xi}$ for individual matrix elements. It is evident that the resulting matrix is positive-definite and therefore contributes trivially (i.e., with zero phase) to the phase factor. The analysis for the bulk of the other three regions follows an identical logic.

Next, we consider contributions from the interface between two regions, localized near their 1D boundary. Let $({BD})_0$ denote such a region, defined by $j\in ({BD})_0$ where $\dist(j, {AC})$ is sufficiently large. Projecting onto this subspace, we obtain:
\begin{equation}
\begin{aligned}
    &\quad~~(\BB\DD)_0M (\BB\DD)_0\\
    &\approx  (\BB\DD)_0\left(\Id-P+P(\ee^{\lambda+\mu}P_B^{\alpha+\beta}+P_D^{\alpha+\beta})P\right) (\BB\DD)_0 \\
  &\approx  (\BB\DD)_0\left(\Id-P+P(\ee^{\lambda+\mu}P_B^{\alpha+\beta}+P_{I-B}^{\alpha+\beta})P\right) (\BB\DD)_0 \\
\end{aligned}
\end{equation}
To analyze the matrix in the final equality, we note that the operator $P_B$ has eigenvalues $x$ satisfying $0\le x\le1$. For small $\lambda$ and $\mu$, the term $\ee^{\lambda+\mu} x^{\alpha+\beta}+(1-x)^{\alpha+\beta}$ remains strictly positive. Therefore, when restricted to $({BD})_0$, $M$ is a positive-definite matrix and does not contribute to the complex phase. Following the same logic, the other 1D interfaces also do not contribute.
\subsubsection{Near the triple contact points}
We are now in a position to evaluate the dominant contributions. Having established that neither the 2D bulk nor the 1D interfaces contribute to the phase, the non-trivial contributions must arise exclusively from the 0D triple contact points. Indeed, in these regions, the restricted matrix $M$ is no longer positive-definite and thus generates a non-zero phase. Focusing on the $ABC$ junction, we can neglect the spatially separated region $D$ and take the thermodynamic limit where $\AA+\BB+\CC \to \Id$:
\begin{equation*}
    \adjincludegraphics[scale=1,valign=c]{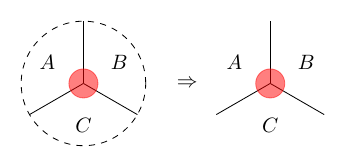}
\end{equation*}
In this limit, the combined region $ABC$ extends to infinity, allowing us to simplify the expression for $M$ as follows:
\begin{equation}
\begin{aligned}
     M_{D=\emptyset}&=\Id -P + P(\ee^\lambda P_{AB}^\alpha+P_{C}^\alpha)(\ee^\mu P_{BC}^\beta+P_{A}^\beta)P \\ &= \mathbb{I}-P +P(\ee^{\lambda+\mu}P_{AB}^\alpha P_{BC}^\beta +\ee^{\mu}P_{C}^\alpha P_{BC}^\beta +\ee^\lambda P_{AB}^\alpha P_A^\beta)P .\\
\end{aligned}
\label{eq:Dempty}
\end{equation}
We can simplify this further by introducing the notation $\widetilde{\RR}\equiv PRP$ and utilizing the identity $(\Id-P)(\Id-\widetilde{R})=\Id-P$. This yields:
\begin{equation}
   M_{D=\emptyset}=(\Id-P+\ee^{\lambda+\mu}P)M_{ABC}(-\lambda,-\mu)
\end{equation}
The first factor, $(\Id-P+\ee^{\lambda+\mu}P)$, is positive-definite for real $\lambda$ and $\mu$, and therefore contributes zero phase. We thus isolate the relevant contribution by defining the matrix $M_{ABC}(\lambda,\mu)$:
\begin{equation}
\begin{aligned}
     M_{ABC}(\lambda,\mu)&=(\Id-\TC)^\alpha(\Id-\TA-\TC)(\Id-\TA)^\beta\\
     &\quad+\ee^\lambda \TC^{\alpha+1}(\Id-\TA)^\beta + \ee^{\mu}(\Id-\TC)^\alpha \TA^{\beta+1}.
\end{aligned}
    \label{eq:def_MABC}
\end{equation}
A similar calculation can be performed for the region near the $ACD$ junction, yielding:
\begin{equation}
    \begin{aligned}
        M_{B=\emptyset}=M_{CDA}(\lambda,\mu)\rightarrow M_{ABC}(\lambda,\mu).
    \end{aligned}
\end{equation}
where we apply the cyclic relabeling ${ CDA}\to {ABC}$, which is justified as it preserves the orientation.
The contributions from the remaining triple contact points are evaluated similarly. The final result for the total phase is:
\begin{equation}
\begin{aligned}
     \arg \det M& =\arg \det M_{ABC}(\lambda,\mu)+\arg \det M_{ABC}(-\lambda,-\mu)\\
     &~~+\arg \det M_{ABC}(\lambda,-\mu)+\arg \det M_{ABC}(-\lambda,\mu).
\end{aligned}
    \label{eq:argdetM}
\end{equation}
Notably, the terms involving $+\lambda$ and $-\lambda$ (or $+\mu$ and $-\mu$) appear in pairs. Consequently, all odd-order terms in the expansion with respect to $\lambda$ or $\mu$ cancel automatically.
\subsection{Evaluation of determinant}
We now proceed to evaluate the phase of the determinant $\det M_{ABC}(\lambda,\mu)$. First, we utilize the Baker-Campbell-Hausdorff (BCH) formula to express the matrix~\eqref{eq:def_MABC} as a single exponential operator:
\begin{equation}
    \begin{aligned}
        M_{ABC}(\lambda,\mu)&=\ee^{O_1}\ee^{O_2}\ee^{O_3}\\
        &=\exp{\left\{\sum_{i=1}^3 O_i+\frac{1}{2}\sum_{i<j,i,j=1}^3[O_i,O_j]+\cdots\right\}}
    \end{aligned}
\end{equation} where 
\begin{equation}
    \begin{aligned}
         &O_1 =\log(\Id-\TC)^\alpha  \\
        &O_2 = \log\left(\Id-\TA-\TC+\ee^\lambda\frac{\TC^{\alpha+1}}{(\Id-\TC)^\alpha}+\ee^\mu \frac{\TA^{\beta+1}}{(\Id-\TA)^\beta} \right)  \\
        &O_3 = \log(\Id-\TA)^\beta. \\
    \end{aligned}
\end{equation}
Consequently, the determinant can be written as:
\begin{equation}
    \det M_{ABC}(\lambda,\mu) = \exp\left\{\sum_{i=1}^3\Tr O_i+\frac{1}{2}\sum_{i<j,i,j=1}^3\Tr[O_i,O_j]\right\}.
\end{equation}
Crucially, we observe that under the trace operation, the BCH expansion terminates effectively at the commutator term. This is because the commutator $[O_i,O_j]$ has finite trace; therefore, the trace of higher-order nested commutators vanishes (i.e., $\Tr[[O_1,O_2], X]=0$ for any bounded operator $X$). 
Now 
\begin{equation}
\begin{aligned}
    &~~~~~\ii\arg\det M_{ABC}(\lambda,\mu)\\
    &=\frac{1}{2}\left(\Tr[O_1,O_2]+\Tr[O_1,O_3]+\Tr[O_2,O_3]\right).
\end{aligned}
\end{equation} 

\subsubsection{Lowest order contributions}
The above trace of commutator can be simplified by a useful identity discovered in Ref.~\cite{gass2025renyilike}: 
\begin{equation}
    \begin{aligned}
       \Tr\left[O_1(\TC),O_2(\TA,\TC)\right]&=\Tr\left[O_1(\TC),O_2(\TA,I-\TA)\right] \\
       &\quad\quad\quad -\Tr\left[O_1(\TC),O_2(0,I-\TA)\right]. \\
    \end{aligned}
\end{equation}
Applying this to our specific operators yields:
\begin{equation}
\begin{aligned}
    &\quad~\Tr\left[O_1(\TC),O_2(\TA,\TC)\right]\\
    &=\Tr\left[\log(\Id-\TC)^\alpha, \log\left( \ee^\lambda(\Id-\TA)^{\alpha+\beta+1}+\ee^\mu \TA^{\alpha+\beta+1}\right)\right]\\
    &\quad\quad-\Tr\left[\log(\Id-\TC)^\alpha,\log\left( \ee^\lambda(\Id-\TA)^{\alpha+1}+\TA^{\alpha+1}\right)\right]\\
    &\quad\quad-\Tr\left[\log(\Id-\TC)^\alpha, \log (\Id-\TA)^{\beta}\right]\\
\end{aligned}
\label{eq:traceO1O2}
\end{equation}
While a closed-form analytical expression for Eq.~\eqref{eq:traceO1O2} is generally elusive, we perform a Taylor expansion in $\lambda$ and $\mu$ up to second order. Each term in the resulting series assumes the form of the left-hand side of Eq.~\eqref{eq:rschern}, which allows for evaluation in terms of $\alpha, \beta, \lambda,$ and $\mu$. Detailed derivations of the relevant integrals and the evaluation of the remaining two commutator terms are provided in Appendix~\ref{Sec:details}.

As noted previously, linear-order coefficients do not contribute to the final result because odd-order terms cancel upon summation over the permutations. Combining the contributions, we obtain:
\begin{equation}
\begin{aligned}
     &\quad~\ii\arg\det M_{ABC}(\lambda,\mu)\\
     &=-\frac{\ii\pi\nu(P)}{48}q(\alpha,\beta)+\frac{\ii\nu(P)}{16\pi}\lambda^2 s(\alpha,\beta)+\frac{\ii\nu(P)}{16\pi}\mu^2 s(\beta,\alpha)\\
     &\quad\quad\quad\quad\quad\quad\quad\quad\quad\quad+{\rm linear ~terms~w.r.t.~\mu~or~\lambda}
\end{aligned}
\end{equation}
Substituting this back into Eq.~\eqref{eq:argdetM}, we finally obtain the relation: 
\begin{equation}
    \begin{aligned}
        &\quad~\ii\arg\det M\\
        &=-\frac{\ii\pi\nu(P)}{12}q(\alpha,\beta)+\frac{\ii\nu(P)}{4\pi}\left(\lambda^2 s(\alpha,\beta)+\mu^2 s(\beta,\alpha)\right),\\
    \end{aligned}
    \label{eq:result}
\end{equation}
This result successfully recovers the established result for free fermions by identifying $\nu(P) = c_{-} = 2\pi\sigma_{xy}$. Remarkably, Eq.~\eqref{eq:result} is exact; higher-order corrections such as $\mathcal{O}(\lambda^4)$ vanish identically (see below).
\subsubsection{Higher order terms} 
It is natural to inquire whether higher-order terms in Eq.~\eqref{eq:result} possess additional physical significance. If non-vanishing topological terms persisted at orders $\mathcal{O}(\lambda^4)$ or higher, they might imply novel formulas for extracting the Hall conductance through non-linear responses. However, we argue that such terms are absent; the lowest non-trivial order ($\lambda^2$) is the unique contributor to the formula.

From the perspective of modular flow, our generating function encodes the response to the dynamics generated by the entanglement Hamiltonian. Specifically, the topological information is captured at the leading non-trivial order, which corresponds to the linear response of the system. Higher-order terms would correspond to non-linear responses, which do not typically characterize the 2D Hall conductance in this framework. 
Developing a rigorous theory for the above claim remains a compelling objective for future research.
Below, we confirm this intuitive picture through explicit calculation in free fermion systems.

We illustrate the absence of higher-order topological contributions by examining the second term of $\Tr [O_1, O_2]$ in Eq.~\eqref{eq:traceO1O2}. By applying the general trace-commutator formula in Eq.~\eqref{eq:rschern}, we identify the relevant functions  as:
\begin{equation}
 f(t)=\log(1-t),\quad g(t)=\log\left(\frac{t^{n} +\ee^{\lambda}(1-t)^{n} }{t^{n}+(1-t)^{n}}\right).
\end{equation}
where we have subtracted the zeroth-order term in $g(t)$.
By introducing the variable change $x=\frac{(1-t)^{n}}{t^{n} +(1-t)^{n}}$, we can expand the function $g(t)$ in powers of $\lambda$:
\begin{equation}
\begin{aligned}
\log\left(\frac{t^{n} +\ee^{\lambda}(1-t)^{n} }{t^{n}+(1-t)^{n}}\right) &= \log\left(1+x(\ee^{\lambda}-1)\right) \\
&=\sum_{k=1}^\infty \frac{\lambda^{k}}{k!}\tilde{g}_{k}(t)=\sum_{k=1}^\infty \frac{\lambda^{k}}{k!}g_{k}(x).
\end{aligned}
\end{equation}
We now focus on the even-order terms $\mathcal{O}(\lambda^{2k})$. The task reduces to analyzing the following integral:
\begin{equation}
\begin{aligned}
 \int_0^1\log(1-t)\partial_t \tilde{g}_{2k}(1-t)\mathrm{d} t &= \int_0^1\frac{\tilde{g}_{2k}(1-t)}{1-t}\mathrm{d} t \\
 &=\frac{1}{2n}\int_0^1 \frac{g_{2k}(x)}{x(1-x)}\mathrm{d} x.
\end{aligned}
\end{equation}
In the first step, we performed integration by parts, and in the second step, we utilized the symmetry property $\tilde{g}_{2k}(t)=\tilde{g}_{2k}(1-t)$ alongside the change of variable to $x$.

For the lowest order case ($k=1$), we have $g_2(x)=x(1-x)$, which straightforwardly reproduces the established $\mathcal{O}(\lambda^2)$ result. For generic $k\ge 2$ (see Appendix~\ref{Sec:details}), $g_{2k}(x)$ becomes a polynomial whose coefficients are expressed in terms of the Stirling numbers of the second kind, denoted as $\StirII{N}{M}$:
\begin{equation}
\frac{g_{2k}(x)}{x(1-x)}=\sum_{m=0}^{2k-2} (-1)^m (m+1)!\StirII{2k-1}{m+1}x^m.
\end{equation}
The integration is now elementary, amounting to the replacement $x^m \to \frac{1}{m+1}$. Remarkably, upon summing the series, the right-hand side vanishes exactly for any $k \ge 2$, a result guaranteed by the combinatorial identities of Stirling numbers. Thus, no higher-order corrections exist.

For the first term of $\Tr[O_1,O_2]$, just noticing that 
\begin{equation}
\log\left(\frac{\ee^\mu t^{n} +\ee^{\lambda}(1-t)^{n} }{t^{n}+(1-t)^{n}}\right) = \log\left(1+x(\ee^{\lambda-\mu}-1)\right)  +\log \ee^\mu,
\end{equation}
then it can be analyzed similarily.
\section{Application}
\label{Sec:app}
\begin{figure}
    \includegraphics[scale=0.5]{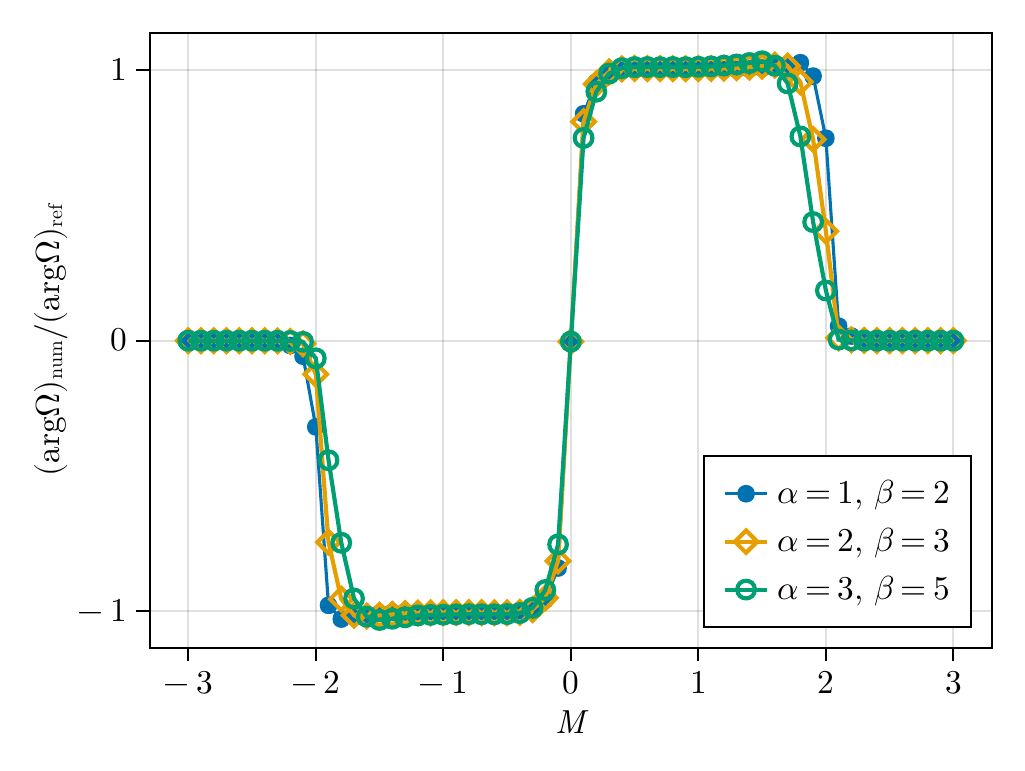}
    \caption{Numerical verification of the generating function $\Omega_{\alpha,\beta,\lambda,\mu}$ for the QWZ model on a $L=32$ lattice with subsystem dimensions $L_{ABC}=16$. With charge parameters fixed at $\lambda=2$ and $\mu=1$, we show the ratio $\arg(\Omega)_{\text{num}} / \arg(\Omega)_{\text{ref}}$ for different replica indices $\alpha$ and $\beta$. The reference phase $\arg(\Omega)_{\text{ref}}$ is obtained from the analytical expression in Eq.~\eqref{eq:result} by setting $c_{-} = 2\pi\sigma_{xy} = 1$. The convergence of the ratio  across different $M$-values confirms the universal scaling predicted by our theory.}
    \label{fig:num}
\end{figure}
In this section, we apply the developed generating function to determine the response of various entanglement measures to modular flow. We further validate the universal form of $\Omega_{\alpha,\beta,\lambda,\mu}$ through numerical simulations on a lattice model.
\subsection{Derivation of the universal response}
We begin by examining the derivative of the R\'{e}nyi  entropy under modular flow. A direct calculation yields:
\begin{equation}
    \left.\frac{\dif{}}{\dif{s}}\right|_{s=0} S_{AB}^{(\alpha)}(s) = \frac{\ii\alpha}{1-\alpha} \frac{\lrangle{[\rho_{AB}^{\alpha-1},K_{BC}]}}{\Tr \rho_{AB}^\alpha}.
    \label{eq:Salpha_ds}
\end{equation}
To evaluate this expression,  we introduce
\begin{equation}
    \Omega_{\alpha\beta}=\lrangle{\rho_{AB}^\alpha\rho_{BC}^\beta}=N_{\alpha\beta}\ee^{-\ii\frac{\pi c_{-}}{12}q(\alpha,\beta)}.
\end{equation}
The term involving the entanglement Hamiltonian $K_{BC}$ can be extracted by taking the derivative with respect to the replica index $\beta$:
\begin{equation}
\begin{aligned}
     \lrangle{[\rho_{AB}^{\alpha-1},K_{BC}]}&=-\left.\frac{\dif{}}{\dif{\beta}}\right|_{\beta=0}(\Omega_{\alpha-1,\beta}-\Omega_{\alpha-1,\beta}^{*}) \\
     &= N_{\alpha-1,0}\ii \frac{\pi c_{-}}{6} \frac{\alpha^2-1}{\alpha^2}
\end{aligned}
\end{equation}
Substituting this result into Eq.~\eqref{eq:Salpha_ds} provides the universal response of the R\'{e}nyi  entropy. A parallel derivation applies to the charged R\'{e}nyi  entropy $S_{AB}^{Q(\alpha)}$, by performing the substitution $\rho_{AB}^{\alpha-1} \to e^{\lambda Q_{AB}} \rho_{AB}^{\alpha-1}$ in Eq.~\eqref{eq:Salpha_ds}.

Finally, we highlight the applicability of our formalism to other measures of quantum correlation. For a modular-flowed pure state $|\psi(s)\rangle_{ABCD}$, the entanglement negativity (and its Rényi generalizations) between subsystems $AB$ and $CD$ can be obtained via analytic continuation of the replica index. Consequently, the response of entanglement negativity to modular flow provides an additional probe for the chiral topological invariant $c_-$.
\subsection{Numerical simulations}
In this subsection, we validate our analytical results through numerical simulations of the 2D Qi-Wu-Zhang (QWZ) model, a prototypical lattice model for a Chern insulator. The Hamiltonian is defined as:
\begin{equation}
    H=\sum_{\vec{k}} \begin{pmatrix}
        a_k^\dagger&b_k^\dagger
    \end{pmatrix}
   \vec{h}(k)\cdot \vec{\sigma}
    \begin{pmatrix}
        a_k \\
        b_k
    \end{pmatrix},
\end{equation}
where the momentum-space vector is $\vec{h}(\vec{k}) = (\sin k_x, \sin k_y, M - \cos k_x - \cos k_y)$ and $\vec{\sigma}$ denotes the vector of Pauli matrices. This model realizes a Chern insulator phase for $|M| < 2$.

We consider a real-space tripartition on a square lattice, similar to Ref.~\cite{sheffer2025probing}. To compute the generating function $\Omega_{\alpha, \beta, \lambda, \mu}$, we utilize the correlation matrix method for free fermions in Eq.~\eqref{eq:def_M}. As shown in Fig.~\ref{fig:num}, the extracted topological invariants match well with thermodynamic limit result.
\section{Summary and discussion}
In this work, we have presented a universal response of entanglement measures to modular flow, with results unified by a single generating function $\Omega_{\alpha,\beta,\lambda,\mu}$. Our results demonstrate that the phase of this quantity encodes the chiral topological invariants, whereas its magnitude is non-universal.

A natural question concerns the well-definedness of this observable. Our analysis in Sec.~\ref{Sec:freefermion} indicates that the underlying matrix $M$ is, at minimum, "locally invertible." Provided global invertibility holds, the quantity remains well-defined. A related technical subtlety involves the convergence of the BCH formula and  isolating the phase from the magnitude. These mathematical issues can be rigorously resolved by introducing a regularization scheme $\Id\to \Id_\epsilon$. As these specific technical challenges were comprehensively addressed in Ref.~\cite{gass2025renyilike}, we refer interested readers to that work for a detailed mathematical treatment.

Several intriguing directions for future research emerge from our findings. First, it would be worthwhile to investigate the response of a broader class of entanglement measures to modular flow. While we have pointed out the pure-state negativity, exploring the mixed-state entanglement negativity of $\rho_{ABC}(s)$ represents a significant next step. Since the state $\rho_{ABC}$ typically becomes mixed after tracing out the environment, characterizing its entanglement dynamics could reveal more nuanced features of the underlying topological phase.

Recent developments have highlighted that chirality in $(2+1)D$ systems is deeply intertwined with tripartite entanglement. Guided by this insight, it is essential to develop new quantities that specifically characterize tripartite correlations. Studying how these quantities respond to modular flow could provide a more refined set of probes for chiral central charges and other topological invariants that are otherwise difficult to isolate.

Broadly, our derivation contributes to the framework of real-space definitions for topological invariants. While numerous real-space observables have been proposed for systems with short- or long-range entanglement, the ultimate objective is to rigorously define these quantities in interacting systems. A promising avenue is to integrate these proposals with the entanglement bootstrap program~\cite{SHI2020} and Kitaev's framework for the local definition of quantum phases~\cite{KitaevLocal}. Such an approach offers a valuable microscopic lattice (UV) perspective on topological quantum field theories.

\emph{Acknowledgments.---}We thank Shenghan Jiang, Guancheng Lu, and Ruihua Fan for useful comments. We are especially grateful to Yingfei Gu for invaluable discussions and feedback on this project.
This work is supported by MOST NO. 2022YFA1403902, NSFC Grant
No.12574179. 
\appendix
\section{Some technical details}
\label{Sec:details}
\subsection{Derivation of $M$}
For exponential operators of the form $\ee^{\hat{X}}$, where $\hat{X}$ is a quadratic form $\hat{X}=\sum_{jk}c_j^\dagger X_{jk}c_k$, the expectation value in a zero-temperature free fermion ground state is given by~\cite{Klich2003}:
\begin{equation}
\lrangle{\ee^{\hat{X}}\ee^{\hat{Y}}}=\det(I-P+P\ee^X \ee^YP).
\end{equation}
In the present context, the charge operator is defined as $\hat{Q}_{R}=\sum_{jk}c_j^\dagger (\RR)_{jk}c_k$, and the reduced density matrix takes the form:
\begin{equation}
 \rho_{R}=\frac{\ee^{-\sum_{jk}c_j^\dagger(k_{R})_{jk}c_k}}{Z_{R}},\quad \text{with } k_{R}=\log \frac{I-P_{R}}{P_{R}}.
\end{equation}
Applying the trace formula, the generating function $\Omega_{\alpha,\beta,\lambda,\mu}$ is evaluated as:
\begin{equation}
\begin{aligned}
&\det\left(I-P+P\ee^{-\alpha k_{AB}}\ee^{\lambda\AA\BB}\ee^{\mu\BB\CC} \ee^{-\beta k_{BC}}P\right)/Z_{AB}^{\alpha}Z_{BC}^\beta.\\
\end{aligned}
\end{equation}
To simplify the expression inside the determinant, we utilize the identity noted in Ref.~\cite{gass2025renyilike}:
\begin{equation}
\begin{aligned}
   & P\ee^{-\alpha k_{AB}}=(I-P\AA\BB P)^{-\alpha} P(P_{AB}^\alpha+P_{CD}^\alpha)\\
   & \ee^{-\alpha k_{BC}}P=(P_{BC}^\beta+P_{AD}^\beta)P(1-P\BB\CC P)^{-\beta}
\end{aligned}
\end{equation}
Finally, the prefactor $(I-P\AA\BB P)^{-\alpha}$ cancels exactly with the normalization constant $Z_{AB}^\alpha = \det(I-P\AA\BB P)^{-\alpha}$, yielding the desired determinant expression for $M$.
\subsection{Trace of commutator}
We show how to evaluate $\Tr[O_1,O_2]$ in Eq.~\eqref{eq:traceO1O2}. We consider the following Taylor expansion up to second orders:
\begin{equation}
    \begin{aligned}
&\log(\ee^{\lambda}X+\ee^{\mu}Y)=\log(X+Y)+\frac{\lambda^2+\mu^2}{2}\frac{XY}{(X+Y)^2}+\cdots\\
    \end{aligned}
\end{equation}
Then we apply each term to Eq.~\eqref{eq:rschern} and get 
\begin{equation}
    \begin{aligned}
       \Tr[O_1,O_2]&=-\frac{\ii \pi\nu(P)}{24}\alpha(\frac{1}{\alpha+1}-\beta-\frac{1}{\alpha+\beta+1}) \\ &+\frac{\ii\nu(P)}{8\pi}(\lambda^2+\mu^2) \frac{\alpha}{\alpha+\beta+1} -\frac{\ii\nu(P)}{8\pi}\lambda^2 \frac{\alpha}{\alpha+1} 
    \end{aligned}
\end{equation}
We could also find (by tracking the orientation)
\begin{equation}
    \Tr[O_2,O_3] = \Tr[O_1,O_2](\alpha\leftrightarrow\beta,\lambda\leftrightarrow\mu)
\end{equation}\
Last, we have 
\begin{equation}
    \Tr[O_1,O_3]=-\frac{\ii\pi\nu(P)}{12}\alpha\beta
\end{equation}
Summarizing the above three terms, we reproduce the main text results.

The relevant integrals are:
\begin{equation}
    \begin{aligned}
        &\int_0^1 \log(1-t)\partial_t \log t \dif t= -\frac{\pi^2}{6} \\
        & \int_0^1 \log(1-t)\partial_t \log\left(t^y+(1-t)^y\right) \dif t = -\frac{\pi^2}{12}(y-\frac{1}{y})\\
        & \int_0^1 \log(1-t)\partial_t  \frac{t^y(1-t)^y}{(t^y+(1-t)^y)^2}\dif t =\frac{1}{2y}.
    \end{aligned}
\end{equation}
\subsection{Expansion of $g(t)$}
We consider the following Taylor expansion:
\begin{equation}
    \begin{aligned}
        \log(1+(\ee^{\lambda}-1)x)&=\sum_{j=1}^\infty (-1)^{j+1}\frac{x^j}{j} (\ee^{\lambda}-1)^j \\
        &=  \sum_{j=1}^\infty (-1)^{j+1}\frac{x^j}{j} \sum_{k=j}^\infty \frac{j!}{k!}\StirII{k}{j}(\lambda)^k \\
        &=\sum_{k=1}^\infty \frac{(\lambda)^k}{k!}\sum_{j=1}^k(-1)^{j-1}(j-1)! \StirII{k}{j}x^j
    \end{aligned}
\end{equation}
The coefficient polynomials are:
\begin{equation}
    g_{2k}(x)=\sum_{j=1}^{2k}a_j x^j,\quad a_j=(-1)^{j-1}(j-1)! \StirII{2k}{j}
\end{equation}
Now $g_{2k}/x(1-x)$ can be obtained by standard polynomial division with coefficients can be simplified by the following identities:
\begin{equation}
\left\{
\begin{aligned}
   & \sum_{r=0}^{2k-1} (-1)^r r! \StirII{2k}{r+1}=0  \\
   & \sum_{r=0}^{m} (-1)^r r! \StirII{2k}{r+1}=(-1)^m (m+1)! \StirII{2k-1}{m+1}
\end{aligned}
\right.
\end{equation}
We then get the final result:
\begin{equation}
\begin{aligned}
     \frac{g_{2k}(x)}{x(1-x)}
     &= \sum_{m=0}^{2k-2} (-1)^m (m+1)!\StirII{2k-1}{m+1}x^m
\end{aligned}
\end{equation}

\section{Field theory method}
\label{Sec:field_theory}
\begin{figure}[t]
\centering
   \adjincludegraphics[scale=1,valign=c]{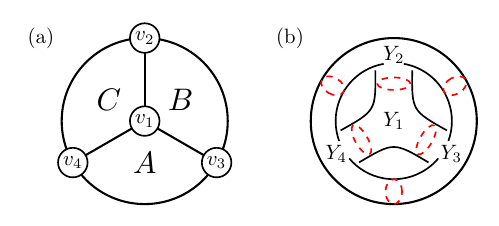}
\caption{(a): Tripartited region $ABC$ with four triple points $v_1,v_2,v_3,v_4$. (b): regularization surface $Y$ is cut into four 3-punctured spheres $Y_v$, $v=1,2,3,4$.}
\label{fig:def_Y}
\end{figure}

\begin{figure}[t]
    \centering
    \adjincludegraphics[scale=1,valign=c]{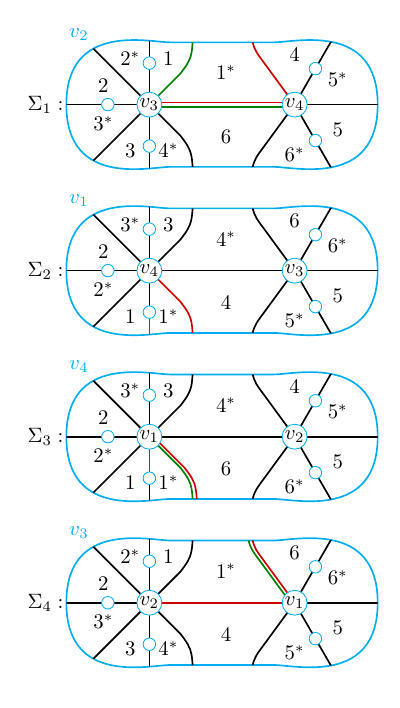}
    \caption{Example for $\Sigma_v$ with $m=3$, $n=2$. Red/green lines denotes $\lambda$/$\mu$-flux. For simplicity, we don't draw geodesic lines. We notice there is an ambiguity in choosing $\pm\lambda$ or $\pm\mu$ in $\theta_{vv^\prime}$, but both choices give the same $\theta_{vv^\prime}^2$ results.
    }
    \label{fig:sigmav}
\end{figure}
In this section, we outline the derivation of the generating function using field-theoretic methods. 
We refer the readers to Ref.~\cite{sheffer2025probing} for more details.
To facilitate comparison with the literature, we adopt the tripartite convention illustrated in Fig.~\ref{fig:def_Y}(a). We begin by expressing the generating function $\Omega$ in its replica form:
\begin{equation}
\begin{aligned}
      &\quad\lrangle{\rho_{AC}^m\ee^{\lambda {Q}_{AC}}\ee^{\mu{Q}_{AB}} \rho_{AB}^n}\\
      &=\bra{\psi^{\otimes m+n+1}} \ee^{\lambda {Q}_{AC}^{(1)}}\ee^{\mu{Q}_{AB}^{(1)}} \pi_A \pi_B \pi_C\ket{\psi^{\otimes m+n+1}}
\end{aligned}
\end{equation}
utilizing a total of $(m+n+1)$ replicas. Here, $\pi_A, \pi_B,$ and $\pi_C$ are permutation operators acting on the respective regions within the replica space. Their cycle structures are defined as:
\begin{equation}
    \begin{aligned}
        &\pi_A = (1,2,\dots,m+n+1) \\
        &\pi_B = (m+1,m+2,\dots,m+n+1)\\
        &\pi_C=(1,2,\dots,m+1)
    \end{aligned}
\end{equation}
The charge operators $Q_{AC}^{(1)}$ and $Q_{AB}^{(1)}$ act exclusively on the first replica. We interpret $\Omega$ as the partition function of a Topological Quantum Field Theory (TQFT) defined on a three-manifold ${M}$, constructed via the branched covering specified by the permutations $\pi_{A/B/C}$. For systems with a non-vanishing chiral central charge $c_-$, the entanglement cuts must be regularized using a chiral CFT. The original entanglement surface ${Y}$ [Fig.~\ref{fig:def_Y}(b)] is cut and reglued according to the permutation rules to form a closed two-manifold $\Sigma$. The phase of the partition function is determined entirely by the physics on $\Sigma$:
\begin{equation}
    \arg \Omega = \arg Z_{{\rm CFT}}(\Sigma).
\end{equation}
The evaluation of the generating function thus reduces to computing the 2D CFT partition function on $\Sigma$, with the geometry and connectivity dictated by $\pi_{A/B/C}$.

Following the prescription in Ref.~\cite{sheffer2025probing}, we decompose the entanglement surface ${Y}$ into four 3-punctured spheres $Y_{v}$ ($v=1,2,3,4$). Gluing $(m+n+1)$ copies of each $Y_v$ results in four surfaces $\Sigma_v$. This initial stage of the construction does not contribute to the complex phase.  The phase arises when we glue the four $\Sigma_v$ surfaces together by connecting their respective holes (punctures). For instance, as shown in Fig.~\ref{fig:sigmav}, we glue $\Sigma_1$ to $\Sigma_2$ by joining their $v_2$ and $v_1$ holes. This process introduces a geometric twist $\tau_{vv^\prime}$ and a $U(1)$-flux $\theta_{vv^\prime}$, contributing to the partition function as:
\begin{equation}
    Z_{{\rm CFT}}(\Sigma)\propto \exp\{-\ii\frac{c_{-}}{24}\sum_{\{vv^\prime\} }\tau_{vv^\prime} +\ii\frac{\sigma_{xy}}{4\pi}\sum_{\{vv^\prime\}}\theta_{vv^\prime}^2\tau_{vv^\prime}\}.
\end{equation}

The values of the twists $\tau_{vv^\prime}$ are determined by the geodesic lines in each $\Sigma_v$~\cite{sheffer2025probing}. The resulting geometric parameters and associated fluxes are (see Fig.~\ref{fig:sigmav} for an example with $m=3,n=2$):
\begin{equation}
    \begin{aligned}
       & \tau_{v_1 v_2} = \tau_{v_3 v_4} = -\pi \frac{m+n}{m+n+1} \\
       &\tau_{v_1 v_3} = \tau_{v_2 v_4} = \pi \frac{m}{m+1} \\
       &\tau_{v_1 v_4} = \tau_{v_2 v_3} = \pi \frac{n}{n+1}\\
       &\theta_{v_1 v_2}=\ii(\lambda-\mu),\quad\theta_{v_3 v_4}=\ii(\lambda+\mu)\\
       &\theta_{v_1 v_3}=\ii\lambda,\quad\theta_{v_2 v_4}=\ii\lambda\\
       &\theta_{v_1 v_4}=\ii\mu,\quad\theta_{v_2v_3}=\ii\mu,
    \end{aligned}
\end{equation}
Substituting these values into the phase expression allows for the direct recovery of the universal terms found in the main text derivation.
\bibliography{bibpaper}

\begin{thebibliography}{22}%
\makeatletter
\providecommand \@ifxundefined [1]{%
 \@ifx{#1\undefined}
}%
\providecommand \@ifnum [1]{%
 \ifnum #1\expandafter \@firstoftwo
 \else \expandafter \@secondoftwo
 \fi
}%
\providecommand \@ifx [1]{%
 \ifx #1\expandafter \@firstoftwo
 \else \expandafter \@secondoftwo
 \fi
}%
\providecommand \natexlab [1]{#1}%
\providecommand \enquote  [1]{``#1''}%
\providecommand \bibnamefont  [1]{#1}%
\providecommand \bibfnamefont [1]{#1}%
\providecommand \citenamefont [1]{#1}%
\providecommand \href@noop [0]{\@secondoftwo}%
\providecommand \href [0]{\begingroup \@sanitize@url \@href}%
\providecommand \@href[1]{\@@startlink{#1}\@@href}%
\providecommand \@@href[1]{\endgroup#1\@@endlink}%
\providecommand \@sanitize@url [0]{\catcode `\\12\catcode `\$12\catcode `\&12\catcode `\#12\catcode `\^12\catcode `\_12\catcode `\%12\relax}%
\providecommand \@@startlink[1]{}%
\providecommand \@@endlink[0]{}%
\providecommand \url  [0]{\begingroup\@sanitize@url \@url }%
\providecommand \@url [1]{\endgroup\@href {#1}{\urlprefix }}%
\providecommand \urlprefix  [0]{URL }%
\providecommand \Eprint [0]{\href }%
\providecommand \doibase [0]{https://doi.org/}%
\providecommand \selectlanguage [0]{\@gobble}%
\providecommand \bibinfo  [0]{\@secondoftwo}%
\providecommand \bibfield  [0]{\@secondoftwo}%
\providecommand \translation [1]{[#1]}%
\providecommand \BibitemOpen [0]{}%
\providecommand \bibitemStop [0]{}%
\providecommand \bibitemNoStop [0]{.\EOS\space}%
\providecommand \EOS [0]{\spacefactor3000\relax}%
\providecommand \BibitemShut  [1]{\csname bibitem#1\endcsname}%
\let\auto@bib@innerbib\@empty
%</preamble>
\bibitem [{\citenamefont {Kitaev}\ and\ \citenamefont {Preskill}(2006)}]{Kitaev2006TEE}%
  \BibitemOpen
  \bibfield  {author} {\bibinfo {author} {\bibfnamefont {A.}~\bibnamefont {Kitaev}}\ and\ \bibinfo {author} {\bibfnamefont {J.}~\bibnamefont {Preskill}},\ }\bibfield  {title} {\bibinfo {title} {Topological entanglement entropy},\ }\href {https://doi.org/10.1103/PhysRevLett.96.110404} {\bibfield  {journal} {\bibinfo  {journal} {Phys. Rev. Lett.}\ }\textbf {\bibinfo {volume} {96}},\ \bibinfo {pages} {110404} (\bibinfo {year} {2006})}\BibitemShut {NoStop}%
\bibitem [{\citenamefont {Levin}\ and\ \citenamefont {Wen}(2006)}]{Levin2006detecting}%
  \BibitemOpen
  \bibfield  {author} {\bibinfo {author} {\bibfnamefont {M.}~\bibnamefont {Levin}}\ and\ \bibinfo {author} {\bibfnamefont {X.-G.}\ \bibnamefont {Wen}},\ }\bibfield  {title} {\bibinfo {title} {Detecting topological order in a ground state wave function},\ }\href {https://doi.org/10.1103/PhysRevLett.96.110405} {\bibfield  {journal} {\bibinfo  {journal} {Phys. Rev. Lett.}\ }\textbf {\bibinfo {volume} {96}},\ \bibinfo {pages} {110405} (\bibinfo {year} {2006})}\BibitemShut {NoStop}%
\bibitem [{Note1()}]{Note1}%
  \BibitemOpen
  \bibinfo {note} {See, for example Ref.~\cite {Tu2013Momentum,Shiozaki2017,Shiozaki2018, Cian2021,liu2024anyon,h53w-64dy}}\BibitemShut {NoStop}%
\bibitem [{\citenamefont {Shi}\ \emph {et~al.}(2020)\citenamefont {Shi}, \citenamefont {Kato},\ and\ \citenamefont {Kim}}]{SHI2020}%
  \BibitemOpen
  \bibfield  {author} {\bibinfo {author} {\bibfnamefont {B.}~\bibnamefont {Shi}}, \bibinfo {author} {\bibfnamefont {K.}~\bibnamefont {Kato}},\ and\ \bibinfo {author} {\bibfnamefont {I.~H.}\ \bibnamefont {Kim}},\ }\bibfield  {title} {\bibinfo {title} {Fusion rules from entanglement},\ }\href {https://doi.org/https://doi.org/10.1016/j.aop.2020.168164} {\bibfield  {journal} {\bibinfo  {journal} {Annals of Physics}\ }\textbf {\bibinfo {volume} {418}},\ \bibinfo {pages} {168164} (\bibinfo {year} {2020})}\BibitemShut {NoStop}%
\bibitem [{\citenamefont {Kitaev}(2025)}]{KitaevLocal}%
  \BibitemOpen
  \bibfield  {author} {\bibinfo {author} {\bibfnamefont {A.}~\bibnamefont {Kitaev}},\ }\href {https://www.youtube.com/watch?v=CGk48aIei2c} {\bibinfo {title} {Local definitions of gapped hamiltonians and topological and invertible states {I}}} (\bibinfo {year} {2025})\BibitemShut {NoStop}%
\bibitem [{\citenamefont {Kim}\ \emph {et~al.}(2022{\natexlab{a}})\citenamefont {Kim}, \citenamefont {Shi}, \citenamefont {Kato},\ and\ \citenamefont {Albert}}]{Kim2022chiral}%
  \BibitemOpen
  \bibfield  {author} {\bibinfo {author} {\bibfnamefont {I.~H.}\ \bibnamefont {Kim}}, \bibinfo {author} {\bibfnamefont {B.}~\bibnamefont {Shi}}, \bibinfo {author} {\bibfnamefont {K.}~\bibnamefont {Kato}},\ and\ \bibinfo {author} {\bibfnamefont {V.~V.}\ \bibnamefont {Albert}},\ }\bibfield  {title} {\bibinfo {title} {Chiral central charge from a single bulk wave function},\ }\href {https://doi.org/10.1103/PhysRevLett.128.176402} {\bibfield  {journal} {\bibinfo  {journal} {Phys. Rev. Lett.}\ }\textbf {\bibinfo {volume} {128}},\ \bibinfo {pages} {176402} (\bibinfo {year} {2022}{\natexlab{a}})}\BibitemShut {NoStop}%
\bibitem [{\citenamefont {Kim}\ \emph {et~al.}(2022{\natexlab{b}})\citenamefont {Kim}, \citenamefont {Shi}, \citenamefont {Kato},\ and\ \citenamefont {Albert}}]{Kim2022modular}%
  \BibitemOpen
  \bibfield  {author} {\bibinfo {author} {\bibfnamefont {I.~H.}\ \bibnamefont {Kim}}, \bibinfo {author} {\bibfnamefont {B.}~\bibnamefont {Shi}}, \bibinfo {author} {\bibfnamefont {K.}~\bibnamefont {Kato}},\ and\ \bibinfo {author} {\bibfnamefont {V.~V.}\ \bibnamefont {Albert}},\ }\bibfield  {title} {\bibinfo {title} {Modular commutator in gapped quantum many-body systems},\ }\href {https://doi.org/10.1103/PhysRevB.106.075147} {\bibfield  {journal} {\bibinfo  {journal} {Phys. Rev. B}\ }\textbf {\bibinfo {volume} {106}},\ \bibinfo {pages} {075147} (\bibinfo {year} {2022}{\natexlab{b}})}\BibitemShut {NoStop}%
\bibitem [{\citenamefont {Fan}(2022)}]{Fan2022from}%
  \BibitemOpen
  \bibfield  {author} {\bibinfo {author} {\bibfnamefont {R.}~\bibnamefont {Fan}},\ }\bibfield  {title} {\bibinfo {title} {From entanglement generated dynamics to the gravitational anomaly and chiral central charge},\ }\href {https://doi.org/10.1103/PhysRevLett.129.260403} {\bibfield  {journal} {\bibinfo  {journal} {Phys. Rev. Lett.}\ }\textbf {\bibinfo {volume} {129}},\ \bibinfo {pages} {260403} (\bibinfo {year} {2022})}\BibitemShut {NoStop}%
\bibitem [{\citenamefont {Fan}\ \emph {et~al.}(2023{\natexlab{a}})\citenamefont {Fan}, \citenamefont {Sahay},\ and\ \citenamefont {Vishwanath}}]{Fan2023Extracting}%
  \BibitemOpen
  \bibfield  {author} {\bibinfo {author} {\bibfnamefont {R.}~\bibnamefont {Fan}}, \bibinfo {author} {\bibfnamefont {R.}~\bibnamefont {Sahay}},\ and\ \bibinfo {author} {\bibfnamefont {A.}~\bibnamefont {Vishwanath}},\ }\bibfield  {title} {\bibinfo {title} {Extracting the quantum hall conductance from a single bulk wave function},\ }\href {https://doi.org/10.1103/PhysRevLett.131.186301} {\bibfield  {journal} {\bibinfo  {journal} {Phys. Rev. Lett.}\ }\textbf {\bibinfo {volume} {131}},\ \bibinfo {pages} {186301} (\bibinfo {year} {2023}{\natexlab{a}})}\BibitemShut {NoStop}%
\bibitem [{\citenamefont {Fan}\ \emph {et~al.}(2023{\natexlab{b}})\citenamefont {Fan}, \citenamefont {Zhang},\ and\ \citenamefont {Gu}}]{Fan2023generalized}%
  \BibitemOpen
  \bibfield  {author} {\bibinfo {author} {\bibfnamefont {R.}~\bibnamefont {Fan}}, \bibinfo {author} {\bibfnamefont {P.}~\bibnamefont {Zhang}},\ and\ \bibinfo {author} {\bibfnamefont {Y.}~\bibnamefont {Gu}},\ }\bibfield  {title} {\bibinfo {title} {{Generalized real-space Chern number formula and entanglement hamiltonian}},\ }\href {https://doi.org/10.21468/SciPostPhys.15.6.249} {\bibfield  {journal} {\bibinfo  {journal} {SciPost Phys.}\ }\textbf {\bibinfo {volume} {15}},\ \bibinfo {pages} {249} (\bibinfo {year} {2023}{\natexlab{b}})}\BibitemShut {NoStop}%
\bibitem [{Note2()}]{Note2}%
  \BibitemOpen
  \bibinfo {note} {This argument implies that the density of entanglement degrees of freedom, $\sigma $, is inversely proportional to its rate of change $v$ under modular flow. Consequently, the product $\sigma v$ remains dimensionless.}\BibitemShut {Stop}%
\bibitem [{\citenamefont {Sheffer}\ \emph {et~al.}(2025{\natexlab{a}})\citenamefont {Sheffer}, \citenamefont {Fan}, \citenamefont {Stern}, \citenamefont {Berg},\ and\ \citenamefont {Ryu}}]{sheffer2025probing}%
  \BibitemOpen
  \bibfield  {author} {\bibinfo {author} {\bibfnamefont {Y.}~\bibnamefont {Sheffer}}, \bibinfo {author} {\bibfnamefont {R.}~\bibnamefont {Fan}}, \bibinfo {author} {\bibfnamefont {A.}~\bibnamefont {Stern}}, \bibinfo {author} {\bibfnamefont {E.}~\bibnamefont {Berg}},\ and\ \bibinfo {author} {\bibfnamefont {S.}~\bibnamefont {Ryu}},\ }\href {https://arxiv.org/abs/2512.04649} {\bibinfo {title} {Probing chiral topological states with permutation defects}} (\bibinfo {year} {2025}{\natexlab{a}}),\ \Eprint {https://arxiv.org/abs/2512.04649} {arXiv:2512.04649 [quant-ph]} \BibitemShut {NoStop}%
\bibitem [{\citenamefont {Gass}\ and\ \citenamefont {Levin}(2025)}]{gass2025renyilike}%
  \BibitemOpen
  \bibfield  {author} {\bibinfo {author} {\bibfnamefont {J.}~\bibnamefont {Gass}}\ and\ \bibinfo {author} {\bibfnamefont {M.}~\bibnamefont {Levin}},\ }\href {https://arxiv.org/abs/2512.20608} {\bibinfo {title} {R\'enyi-like entanglement probe of the chiral central charge}} (\bibinfo {year} {2025}),\ \Eprint {https://arxiv.org/abs/2512.20608} {arXiv:2512.20608 [cond-mat.str-el]} \BibitemShut {NoStop}%
\bibitem [{\citenamefont {Peschel}(2003)}]{Peschel2003}%
  \BibitemOpen
  \bibfield  {author} {\bibinfo {author} {\bibfnamefont {I.}~\bibnamefont {Peschel}},\ }\bibfield  {title} {\bibinfo {title} {Calculation of reduced density matrices from correlation functions},\ }\href {https://doi.org/10.1088/0305-4470/36/14/101} {\bibfield  {journal} {\bibinfo  {journal} {Journal of Physics A: Mathematical and General}\ }\textbf {\bibinfo {volume} {36}},\ \bibinfo {pages} {L205} (\bibinfo {year} {2003})}\BibitemShut {NoStop}%
\bibitem [{\citenamefont {Klich}(2003)}]{Klich2003}%
  \BibitemOpen
  \bibfield  {author} {\bibinfo {author} {\bibfnamefont {I.}~\bibnamefont {Klich}},\ }\bibinfo {title} {An elementary derivation of levitov's formula},\ in\ \href@noop {} {\emph {\bibinfo {booktitle} {Quantum Noise in Mesoscopic Physics}}},\ \bibinfo {editor} {edited by\ \bibinfo {editor} {\bibfnamefont {Y.~V.}\ \bibnamefont {Nazarov}}}\ (\bibinfo  {publisher} {Springer Netherlands},\ \bibinfo {address} {Dordrecht},\ \bibinfo {year} {2003})\ pp.\ \bibinfo {pages} {397--402}\BibitemShut {NoStop}%
\bibitem [{\citenamefont {Kitaev}(2006)}]{KITAEV2006}%
  \BibitemOpen
  \bibfield  {author} {\bibinfo {author} {\bibfnamefont {A.}~\bibnamefont {Kitaev}},\ }\bibfield  {title} {\bibinfo {title} {Anyons in an exactly solved model and beyond},\ }\href {https://doi.org/https://doi.org/10.1016/j.aop.2005.10.005} {\bibfield  {journal} {\bibinfo  {journal} {Annals of Physics}\ }\textbf {\bibinfo {volume} {321}},\ \bibinfo {pages} {2} (\bibinfo {year} {2006})},\ \bibinfo {note} {january Special Issue}\BibitemShut {NoStop}%
\bibitem [{\citenamefont {Tu}\ \emph {et~al.}(2013)\citenamefont {Tu}, \citenamefont {Zhang},\ and\ \citenamefont {Qi}}]{Tu2013Momentum}%
  \BibitemOpen
  \bibfield  {author} {\bibinfo {author} {\bibfnamefont {H.-H.}\ \bibnamefont {Tu}}, \bibinfo {author} {\bibfnamefont {Y.}~\bibnamefont {Zhang}},\ and\ \bibinfo {author} {\bibfnamefont {X.-L.}\ \bibnamefont {Qi}},\ }\bibfield  {title} {\bibinfo {title} {Momentum polarization: An entanglement measure of topological spin and chiral central charge},\ }\href {https://doi.org/10.1103/PhysRevB.88.195412} {\bibfield  {journal} {\bibinfo  {journal} {Phys. Rev. B}\ }\textbf {\bibinfo {volume} {88}},\ \bibinfo {pages} {195412} (\bibinfo {year} {2013})}\BibitemShut {NoStop}%
\bibitem [{\citenamefont {Shiozaki}\ \emph {et~al.}(2017)\citenamefont {Shiozaki}, \citenamefont {Shapourian},\ and\ \citenamefont {Ryu}}]{Shiozaki2017}%
  \BibitemOpen
  \bibfield  {author} {\bibinfo {author} {\bibfnamefont {K.}~\bibnamefont {Shiozaki}}, \bibinfo {author} {\bibfnamefont {H.}~\bibnamefont {Shapourian}},\ and\ \bibinfo {author} {\bibfnamefont {S.}~\bibnamefont {Ryu}},\ }\bibfield  {title} {\bibinfo {title} {Many-body topological invariants in fermionic symmetry-protected topological phases: Cases of point group symmetries},\ }\href {https://doi.org/10.1103/PhysRevB.95.205139} {\bibfield  {journal} {\bibinfo  {journal} {Phys. Rev. B}\ }\textbf {\bibinfo {volume} {95}},\ \bibinfo {pages} {205139} (\bibinfo {year} {2017})}\BibitemShut {NoStop}%
\bibitem [{\citenamefont {Shiozaki}\ \emph {et~al.}(2018)\citenamefont {Shiozaki}, \citenamefont {Shapourian}, \citenamefont {Gomi},\ and\ \citenamefont {Ryu}}]{Shiozaki2018}%
  \BibitemOpen
  \bibfield  {author} {\bibinfo {author} {\bibfnamefont {K.}~\bibnamefont {Shiozaki}}, \bibinfo {author} {\bibfnamefont {H.}~\bibnamefont {Shapourian}}, \bibinfo {author} {\bibfnamefont {K.}~\bibnamefont {Gomi}},\ and\ \bibinfo {author} {\bibfnamefont {S.}~\bibnamefont {Ryu}},\ }\bibfield  {title} {\bibinfo {title} {Many-body topological invariants for fermionic short-range entangled topological phases protected by antiunitary symmetries},\ }\href {https://doi.org/10.1103/PhysRevB.98.035151} {\bibfield  {journal} {\bibinfo  {journal} {Phys. Rev. B}\ }\textbf {\bibinfo {volume} {98}},\ \bibinfo {pages} {035151} (\bibinfo {year} {2018})}\BibitemShut {NoStop}%
\bibitem [{\citenamefont {Dehghani}\ \emph {et~al.}(2021)\citenamefont {Dehghani}, \citenamefont {Cian}, \citenamefont {Hafezi},\ and\ \citenamefont {Barkeshli}}]{Cian2021}%
  \BibitemOpen
  \bibfield  {author} {\bibinfo {author} {\bibfnamefont {H.}~\bibnamefont {Dehghani}}, \bibinfo {author} {\bibfnamefont {Z.-P.}\ \bibnamefont {Cian}}, \bibinfo {author} {\bibfnamefont {M.}~\bibnamefont {Hafezi}},\ and\ \bibinfo {author} {\bibfnamefont {M.}~\bibnamefont {Barkeshli}},\ }\bibfield  {title} {\bibinfo {title} {Extraction of the many-body chern number from a single wave function},\ }\href {https://doi.org/10.1103/PhysRevB.103.075102} {\bibfield  {journal} {\bibinfo  {journal} {Phys. Rev. B}\ }\textbf {\bibinfo {volume} {103}},\ \bibinfo {pages} {075102} (\bibinfo {year} {2021})}\BibitemShut {NoStop}%
\bibitem [{\citenamefont {Liu}(2024)}]{liu2024anyon}%
  \BibitemOpen
  \bibfield  {author} {\bibinfo {author} {\bibfnamefont {S.}~\bibnamefont {Liu}},\ }\bibfield  {title} {\bibinfo {title} {Anyon quantum dimensions from an arbitrary ground state wave function},\ }\href {https://www.nature.com/articles/s41467-024-47856-7} {\bibfield  {journal} {\bibinfo  {journal} {Nature Communications}\ }\textbf {\bibinfo {volume} {15}},\ \bibinfo {pages} {5134} (\bibinfo {year} {2024})}\BibitemShut {NoStop}%
\bibitem [{\citenamefont {Sheffer}\ \emph {et~al.}(2025{\natexlab{b}})\citenamefont {Sheffer}, \citenamefont {Stern},\ and\ \citenamefont {Berg}}]{h53w-64dy}%
  \BibitemOpen
  \bibfield  {author} {\bibinfo {author} {\bibfnamefont {Y.}~\bibnamefont {Sheffer}}, \bibinfo {author} {\bibfnamefont {A.}~\bibnamefont {Stern}},\ and\ \bibinfo {author} {\bibfnamefont {E.}~\bibnamefont {Berg}},\ }\bibfield  {title} {\bibinfo {title} {Extracting topological spins from bulk multipartite entanglement},\ }\href {https://doi.org/10.1103/h53w-64dy} {\bibfield  {journal} {\bibinfo  {journal} {Phys. Rev. Lett.}\ }\textbf {\bibinfo {volume} {135}},\ \bibinfo {pages} {086601} (\bibinfo {year} {2025}{\natexlab{b}})}\BibitemShut {NoStop}%
\end{thebibliography}%

\end{document}